\documentclass[epjST]{svjour}
\usepackage{epsfig}
\usepackage{graphics}
\usepackage{amsmath}
\usepackage{amssymb}
\newcommand{\be}{\begin{eqnarray}}
\newcommand{\ee}{\end{eqnarray}}
\newcommand{\nee}{\nonumber\end{eqnarray}}
\newcommand{\nn}{\nonumber\\}
\def\a{\alpha}

\begin{document}
\title{On model independent extraction
     of the Kaon Fragmentation Functions}
\author{S. Albino\inst{1}, E. Christova\inst{2}\fnmsep\thanks{\email{echristo@inrne.bas.bg}}  and E. Leader\inst{3}  }
\institute{
Institut f\"ur
Theoretische Physik, Universit\"at Hamburg, Hamburg, Germany\and
Institute for Nuclear Research and Nuclear Energy of
Bulgarian Academy of Sciences, Sofia, Bulgaria\and  Imperial
College, London University, London, UK   }
\abstract{We present a model independent approach to FFs, which allows to
determine uniquely  the  non-singlet combination
$(D_u-D_d)^{K^++K^-}$ of the kaon FFs,  independently in the three
different processes of kaon inclusive production:  $e^+e^-$
annihilation, $pp$-collisions and $lN$ semi-inclusive DIS.
 Only C and SU(2) invariances are used and no
  assumptions on FFs and PDFs, the result holds in any order in
  QCD. The non-singlets obtained in this
 way, would allow to test the commonly used assumptions in global fit analysis and
to  test universality.
This method is applied to   kaon production data in $e^++e^-\to K
+X,\,K=K^\pm, K^0_s$  and  $(D_u-D_d)^{K^++K^-}$ is obtained without
any assumptions. This result is compared to the existing
parametrizations, obtained in global fit analysis.}
%
\maketitle
\section{Introduction}\label{intro}

In order to extract correct information about the
quark and lepton interactions from high energy experiments, the
precise knowledge of two basic quantities are required -- the parton
distribution functions (PDFs), that describe the distribution of the
nucleon momentum among its constituents -- the quarks and gluons,
and the fragmentation functions (FFs), that describe the
distribution of the parton momentum among the hadrons, which the
parton fragments into. Both the PDFs and the FFs are
non-perturbative objects, to be inferred from experiment. Quantum
Chromodynamics (QCD) does not provide a definite picture for their
calculation, it determines only the order by order calculation in
perturbation theory for their energy scale dependence,
described by the $Q^2$-evolution DGLAP equations.

While PDFs are relatively well determined, there are lot of
uncertainties about the FFs at present. The importance of the FFs
has become evident only the last decade, with the start of the new
generation of high-energy experiments with a final hadron $h$ detected
like semi-inclusive deep inelastic scattering (SIDIS) and hadron
production in $pp$-collisions. The above information becomes still
more viable for the LHC, where detecting different final hadrons
will be our  window for the New Physics expected.

At present different parametrizations about the FFs, obtained from
global fit to data,  exist~\cite{DSS,HKNS,AKK08,EMC}. Three characteristic features about them
hold: 1) they use different theoretical assumptions about unfavoured
FFs,  2) they describe the data and 3) they don't agree among
themselves.

Recently, the problem with the kaon FFs has become most appealing
 when  the COMPASS collaboration~\cite{COMPASS} presented two very different
numbers for the strange-quark polarization $\Delta s = \int dx\,\Delta s(x,Q^2)$
 using the same data but two different
parametrizations for the FFs: DSS~\cite{DSS} and EMC~\cite{EMC}:
\be
\Delta s = -0.01\, (DSS),\qquad  \Delta s = -0.04\, (EMC).
\ee

\section{The difference cross sections} \label{sec:1}

\subsection{Cross sections with $K^\pm$ and $K^0_s$}
We consider charged and neutral kaon production in the three semi-inclusive processes:
\be
e^++ e^-&\to&  K+X\label{e+e-}\\
l+N&\to&  l+K+X\label{eN}\\
p+p&\to&  K+X\label{pp}
\ee
where $K$ stands for either charged or neutral kaons, $K=K^+,K^-,K^0_s$.
Note that measuring neutral kaons does not introduce new FFs in these cross sections as
 SU(2) invariance of strong interactions relates neutral and charged kaon FFs:
\be
&&D_u^{K^++K^-}=D_d^{K^0 + \bar K^0},\quad D_d^{K^++K^-}=D_u^{K^0 + \bar K^0},\quad  D_s^{K^++K^-}=D_s^{K^0 + \bar K^0},\nn
 && D_g^{K^++K^-}=D_g^{K^0 + \bar K^0},\quad D_c^{K^++K^-}=D_c^{K^0 + \bar K^0},\quad D_b^{K^++K^-}=D_b^{K^0 + \bar K^0}.\label{SU2}
\ee
We show that the measuring the difference of charged and neutral kaons:
\be
\sigma^{\cal K}=\sigma^{K^++K^--2K^0_s}\equiv \sigma^{K^+}+\sigma^{K^-}-2\sigma^{K^0_s}
\ee
in any of the three processes (\ref{e+e-}) - (\ref{pp}),  always  measures the same NS combination
$(D_u-D_d)^{K^++K^-}$.
This result is obtained without any assumptions about the PDFs and FFs,
using only the SU(2) relations (\ref{SU2}) for the kaons
 and, being a consequence of a symmetry, holds in any QCD order. The results for the three processes are~\cite{FFs}:

 1) For the  $z$-distribution in $e^+e^-\to(\gamma ,Z)\to K+X$ we have:
\be
 d\sigma_{e^+e^-}^{K^++K^--2K_s^0}(z,Q^2)=6\,\sigma_0\, (\hat e_u^2-\hat e_d^2)(1+\frac{\a_s}{2\pi} \, C_q\,\otimes \,)
 \,D_{u-d}^{K^++K^-}(z, Q^2)\label{e+e-K}
 \ee
 where $ \sigma_0=4\pi\alpha_{em}^2/3\,s$ and $\hat e_q$ are the electroweak charges of the quark $q$,
 $C_q$ are the perturbatively calculable Wilson coefficients, $z$ is the fraction of the  momentum of
the fragmenting parton transferred to the hadron $h$: $z=
E^h/E $, where $E^h$ and $E$ are the CM energies of
the final hadron  and the initial lepton, and $\sqrt s=2E$.

2) The considered difference cross-sections (\ref{eN}) for SIDIS  on
proton and deutron targets are given by:
 \be
d\sigma_p^{K^++K^--2K_s^0}&=&\frac{1}{9}[(4\tilde u-\tilde
d)\otimes  (1+\frac{\a_s}{2\pi}C_{qq}) + \frac{\a_s}{2\pi}g\otimes
C_{gq}]\otimes
\,D_{u-d}^{K^++K^-}\label{SIDISpK}\\
d\sigma_d^{K^++K^--2K_s^0}&=&\frac{1}{3} [(\tilde u+\tilde d)\otimes
(1+\frac{\a_s}{2\pi}C_{qq} ) + 2\frac{\a_s}{2\pi} \,g\otimes
C_{gq}]\otimes \,D_{u-d}^{K^++K^-}.\label{SIDISdK}
\ee
Here $\tilde q\equiv u+\bar u$ and $C_{ab}$ are the  perturbatively calculable Wilson coefficients.

3) For $pp\to K+X$, $K=K^\pm, K^0_s$ we obtain:
\be
E^K\frac{d\sigma_{pp}^{K^++K^--2K_s^0}}{d^3P^K}&=&\frac{1}{\pi}
\sum_{a,b}\int dx_a\int dx_b\int \frac{dz}{z}\,
f_a(x_a)f_b(x_b)\,\nn
&&\times \left(d\hat\sigma_{ab}^{uX}+d\hat\sigma_{ab}^{\bar uX}-
d\hat\sigma_{ab}^{dX}-d\hat\sigma_{ab}^{\bar dX}\right)\,D_{u-d}^{K^++K^-}.\label{ppK}
\ee
Here the sum over $a,b$ is over all partons that contribute, $f_{a,b}$ are the PDFs, and
$d\hat\sigma_{ab}^{cX}(s,t,u)$ are  the partonic cross sections for the inclusive processes $a+b\to c+X$ that contribute.
They are functions of the corresponding Mandelstam variables:
\be
s&=&(p_a+p_b)^2,\quad t=(p_a-p_c)^2,\quad u=(p_b-p_c)^2\nn
p_a&=&x_aP_A,\qquad p_b=x_bP_B,\qquad p_c=P^K/z.
\ee
 (In NLO $s,\,t$ and $u$ are independent variables.) In addition, that the expression
for the difference cross section $\sigma^{\cal K}_{pp}$ is considerably simpler than the corresponding
cross sections for single kaon production --
 in NLO  it is only 8 partonic cross sections that
 contribute to $\sigma^{\cal K}_{pp}$, versus 21 to $\sigma^{K^+}_{pp}$.

 Eqs. ~(\ref{e+e-K})--(\ref{ppK}) are four independent measurements
 that determine uniquely the NS combination  of the kaon  FF $D_{u-d}^{K^++K^-}$.
 As $D_{u-d}^{K^++K^-}$ is a NS, no new FFs will enter through its $Q^2$ evolution.
 As these expressions use only SU(2) invariance,  the
 verification of any of them might serve as a test for SU(2) invariance of the kaon FFs, used in all analysis.

In addition, eqs.~(\ref{e+e-K})- (\ref{SIDISdK}) allow  to compare
the NS obtained in $e^+e^-$ at rather high $Q^2\simeq m_Z^2$,
 with those from SIDIS at quite low $Q^2$. This would provide a challenging
test of the $Q^2$-evolution and universality of the FFs.

 As these expressions are  model independent,
 it would be interesting to compare the resulting NS  to the NS obtained with the existing parametrizations
 extracted from
$e^+e^-$ data, obtained with various assumptions. This we shall do in  Section~\ref{sec:3}.

\subsection{Cross sections with $K^\pm$}
Let's consider  charged kaon production in:
\be
l+N&\to& l+K+X\label{eN+}\\
p+p &\to& K+X,\label{pp+}
\ee
  $K=K^+,K^-$. We show that, using only $C$-invariance, the difference
 cross sections $\sigma^{K^+-K^-}\equiv\sigma^{K^+}-\sigma^{K^-}$ in (\ref{eN+}) and (\ref{pp+})
 determine uniquely the NSs $D_u^{K^+-K^-}$ and $D_d^{K^+-K^-}$
 without any  assumptions. The expressions in NLO are~\cite{FFs}:
 \be
&& \sigma_p^{K^+-K^-}\simeq [4u_V \otimes D_u^{K^+-K^-}+d_V \otimes
D_d^{K^+-K^-}]
 \otimes\, (1+\a_s C_{qq}),\label{diffep}\\
&&\sigma_d^{K^+-K^-}\simeq [(u_V+d_V)\otimes (4 D_u+D_d)^{K^+-K^-}]
\otimes\, (1+\a_s C_{qq})\\
&& \sigma_{pp}^{K^+-K^-}\simeq [ L_u\otimes u_V\otimes D_u +
L_d\otimes  d_V\otimes D_d]^{K^+-K^-}.\label{diffpp}
 \ee
Here $L_q$ are known functions of the unpolarized PDFs and partonic cross sections:
\be
&&L_u(x,t,u)= \tilde u(x)\,d\hat\Sigma(s,t,u)
+[\tilde d(x)+\tilde s(x)]\, d\hat\sigma_{qq'}^{qX}(s,t,u)+ g(x)\, d\hat\sigma_{qg}^{(q-\bar q)X}(s,t ,u)\nn
&&L_d(x,t,u)= \tilde
d(x)\,d\hat\Sigma(s,t,u)+[\tilde u(x)+\tilde s(x)]\,
 d\hat\sigma_{qq'}^{qX}(s,t ,u)+ g(x)\, d\hat\sigma_{qg}^{(q-\bar q)X}(s,t ,u)\nn
&&L_s(x,t,u)= \tilde
s(x)\,d\hat\Sigma(s,t,u)+[\tilde u(x)+\tilde d(x)]\,
d\hat\sigma_{qq'}^{qX}(s,t ,u)+ g(x)\, d\hat\sigma_{qg}^{(q-\bar
q)X}(s,t ,u)
\nee
where $d\hat\Sigma$ is the combination:
 \be
 d\hat\Sigma\equiv
\left[\,d\hat\sigma_{qq}^{qX}(s,t ,u) +
\frac{1}{2}\,d\hat\sigma_{q\bar q}^{(q-\bar q)X}(s,t
,u)\right],\qquad \tilde q \equiv q+\bar q.
\ee
 If we assume $D_d^{K^+-K^-}=0$, used in all global analysis, from (\ref{diffep} - (\ref{diffpp}) we obtain:
\be
  \sigma_p^{K^+-K^-}&=& \frac{4}{9}\,u_V  \otimes\, (1+\a_s\, C_{qq})\otimes D_u^{K^+-K^-},\label{DKep}\\
\sigma_d^{K^+-K^-}&=& \frac{4}{9}\,(u_V+d_V)
\otimes\, (1+\a_s \,C_{qq})\otimes   D_u^{K^+-K^-}\\
 \sigma_{pp}^{K^+-K^-}&=&\frac{1}{\pi}\int dx_adx_b\frac{dz}{z}\left[ L_u (x_b,t,u)\otimes u_V(x_a) +
L_u (x_a,t,u)\otimes u_V(x_b\right)] \otimes D_u^{K^+-K^-}.\nn\label{DKpp}
 \ee
Eqs. (\ref{DKep}) - (\ref{DKpp}) present 3 independent measurements for the NS $D_u^{K^+-K^-}$,
  obtained with the only assumption $D_d^{K^+-K^-} = 0$ and all $\sigma^{K^+-K^-}$ are fitted with just  one FF.
As $D_d^{K^+-K^-} = 0$ has been the only assumption, each of these equations would be a test of just this assumption.

\section{$D_{u-d}^{K^++K^-}$ from $e^+e^-$ data}
\label{sec:3}

On basis of eq. (\ref{e+e-K}), analyzing the available data
on $K^\pm$ and $K^0_s$-production in $e^+e^-$-annihilation, we extracted for
the first time  $D_{u-d}^{K^++K^-}$ without any correlations to other FFs.
 There are
several arguments for choosing $e^+e^-$ data: 1) these are the most accurate and  precise data,
2) the NS component doesn't contain unresummed soft gluons divergences,
which allows to use lower values of $z$ ($z>0.001$) than in global fit analysis
($z>0.1$, DSS~\cite{DSS}; $z>0.05$, HKNS~\cite{HKNS} and AKK08~\cite{AKK08}),
3) for the NS component a next-next to leading order (NNLO) fit is
 possible as both the splitting and coefficient functions are known to NNLO, while in global fit analysis only
 NLO calculations are available at present.

The data used in our analysis are in the energy range $\sqrt s = 12 - 189$ GeV, but
  not all data appear equally important.
Eq.(\ref{e+e-K}) implies that the sensitivity of $\sigma^{\cal K}$ to
 $D_{u-d}^{K^++K^-}$ is determined by the $s$-dependence of
 $(\hat e_u^2-\hat e_d^2)(s)$. This difference is the biggest away from the $Z$-pole, i.e. most important for
 our studies are the data with small $s$, $\sqrt s \leq 60$ GeV and big $s$, $\sqrt s\geq 110$ GeV.  Unfortunately,
   the data around the $Z$-pole, where the most abundant and precise data exist, almost does not contribute.

 Our analysis would  be easy if we had data on $K^\pm$ and $K^0_s$ production at {\it the same} $z$ and $s$.
 However, as this is not the case,  we proceed in three steps~\cite{NSs}:
 1) we combine the measurements into 7 energy intervals:
 $\sqrt s$ = 12-14.8; 21.5-22; 29-35; 42.6-44; 58; 91.2; 183-189 GeV,
 2) in each interval we fit $K^0_s$ data {\it purely phenomenologically} and  3) we
parametrize $D_{u-d}^{K^++K^-}$ at a factorization scale $\mu_0 =\sqrt 2$ GeV:
\be
D_{u-d}^{K^++K^-}(z,\mu_0^2)=nz^\a(1-z)^b+n'z^{\a'}(1-z)b'\label{NS}
\ee
and  fit {\it simultaneously}  the parameters in the parametrizations for
$\sigma^{K_s^0}$ and $D_{u-d}^{K^++K^-}$ in a combined analysis of the data on $K^\pm$ and $K^0_s$.
In our analysis we keep the hadron mass $m_h$ as a fitting parameter.

  Small $z$-values imply small hadron energies, which may become of the order or smaller than the mass of the hadron.
  That's why including the small-$z$ and/or small-$s$ data the effects of the hadron mass, i.e. $m_h\neq 0$,
  become significant and have to be taken into account.
In our analysis we must distinguish kinematically between  the
experimentally measured fractions of the final hadron energy $E_h$
and the modulus of its momentum $\vert {\bf p}_h\vert$ to $\sqrt s$:
\be x_E=\frac{E_h}{2\sqrt
s}=x\left(1+\frac{m_h^2}{x^2\,s}\right),\qquad x_p=\frac{\vert{\bf
p}_h\vert}{2\sqrt s}=x\left(1-\frac{m_h^2}{x^2\,s}\right). \ee These
quantities are equal  and  equal to the light cone scaling variable
$x$, $x_E=x_p=x$, only at $m_h= 0$, $x=z$ in LO in QCD.

\begin{figure}
\centerline{\ \ \ \ \ \ \ \includegraphics[width=60mm,angle=-90]{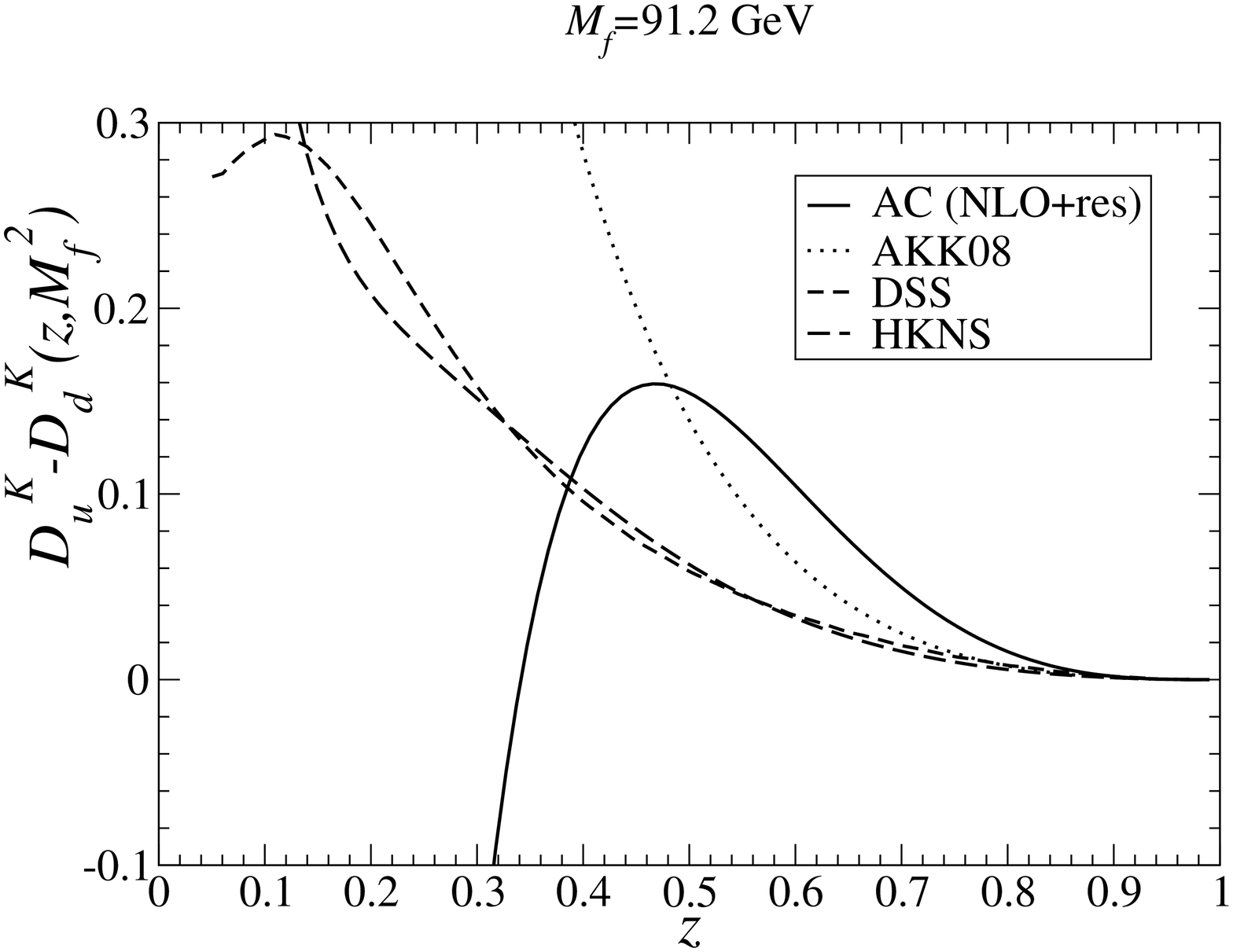}  }
\centerline{\ \ \ \ \ \ \ \includegraphics[width=60mm,angle=-90]{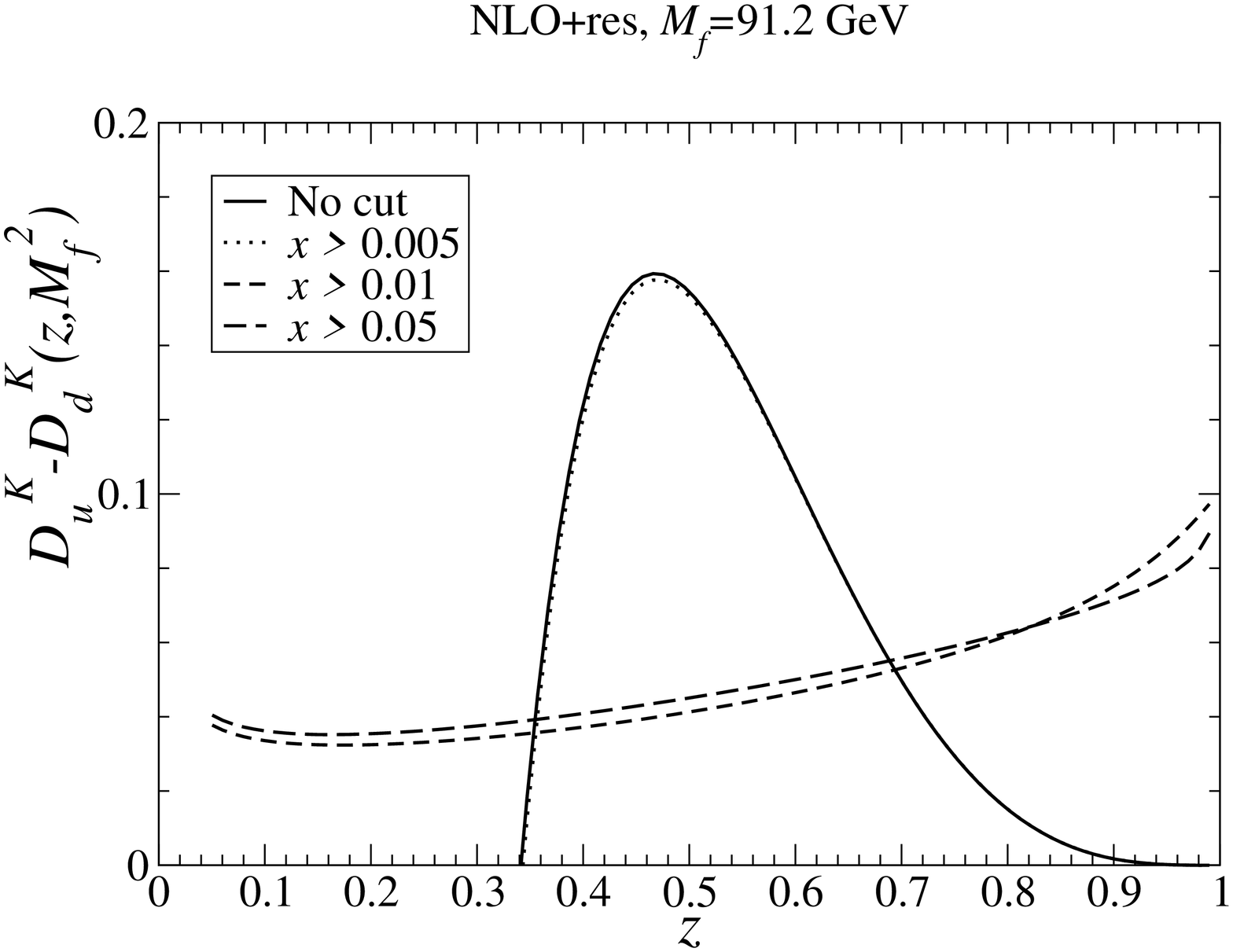} }
\caption{up:  $D_{u-d}^{K^++K^-}$  in this paper at NLO with resumed logarithms [label AC],
 and from  DSS, HKNS and AKK08  sets;
down: the  NS  in this paper with different cuts on the data}
\label{fig:1}       
\end{figure}

The NS $D_{u-d}^{K^++K^-}$  is presented on Fig. \ref{fig:1}, left. On the same figure
$D_{u-d}^{K^++K^-}$, obtained from  the global fits of DSS, AKK08 and HKNS, are presented as well.
The   differences between our
result for $D_{u-d}^{K^++K^-}$ and those of global fit analysis might be due mainly to
1) the different assumptions -- in DSS and HKNS
$D_s^{K^+}=D_{\bar u}^{K^+}=D_d^{K^+}=D_{\bar d}^{K^+}$ is assumed; in AKK08 $D_d^{K^+}=D_{\bar d}^{K^+}$ is used;
we make no assumptions;
2)  the  small $z$-data, for the first time included. In order to get insight on the effect of the low-$z$ data,
on Fig \ref{fig:1}, right we present $D_{u-d}^{K^++K^-}$
  with different cuts on the lower values of $z$.
The result shows that  data at $z\geq 0.01$ is  not enough to impose constraints on the NS.

At the end we point to a striking feature of our result -- the negative value for $D_{u-d}^{K^++K^-}$ at about $z\lesssim 0.3$.
In order to check this result and find out its origin we performed our analysis with
different parametrizations for $D_{u-d}^{K^++K^-}$ and
this does not seem a genuine feature of the NS -- choosing
 $D_{u-d}^{K^++K^-} = nz^\a(1-z)^b$ in stead of (\ref{NS}) yields $D_{u-d}^{K^++K^-}>0$,
  $D_{u-d}^{K^++K^-}=0$ appears also compatible with the data. The quality of the fit only slightly worsens:
 from $\chi^2_{dof}=\chi^2/N=2.2$ for $D_{u-d}^{K^++K^-}<0$, to
 $\chi^2_{dof} \simeq 2.4$ for positive and zero values of the NS,  $N=730$ is the number of points in the fit.
   However,  in all cases the quality of the fit
 is  worse than the one obtained in the purely phenomenological fit to $K^0_s$ - $\chi^2_{dof}=1.1$.

 Also the value of the fitted kaon mass, obtained in the purely phenomenological fit to $K^0_s$ -- $m_K=320$ MeV,
 differs significantly
 from the one obtained in the combined $K^0_s$ and $K^\pm$ analysis based on the QCD formula (\ref{e+e-K}) -- $m_K=124$ MeV.
(The true kaon mass is $m_K$ = 494 MeV.)
A possible reason for this big difference might be that  other low-$\sqrt s$ and small-$x$ effects,
 such as higher twists and mass effects of resonances from which
 kaons may be produced, that have not been accounted for in our analysis, can also be absorbed
in the fitted value for $m_K$.
The performed analysis are very sensitive to hadron mass effects  and
these effects strongly affect our fits.
More accurate data will be needed to determine how important these other effects are.

  Our analysis shows that the non-singlet $D_{u-d}^K$ can be determined uniquely in a combined fit to $e^+e^-$ data,
  but more precise data is needed to fix  it better.

\subsection*{Acknowledgements}

The work of E.\ C.\ was supported by
the Bulgarian Science Foundation, Grant 288/2008 and a collaborative Grant of INRNE (Bulgaria) - JINR (Dubna, Russia).

\end{document}